\documentclass[11pt]{article}
\usepackage{moriond,epsfig}

\def\Journal#1#2#3#4{{#1} {\bf #2}, #3 (#4)}

\def\NIMA{{\it Nucl. Instrum. Methods} A}

\def\be{\begin{equation}}
\def\ee{\end{equation}}
\def\bea{\begin{eqnarray}}
\def\eea{\end{eqnarray}}

\begin{document}
\vspace*{4cm}
\title{CODED-MASK IMAGING IN GAMMA-RAY ASTRONOMY \\
-- Separating the Real and Imaginary parts of a Complex subject}

\author{ G.K. SKINNER }

\address{CESR, 9, avenue du Colonel Roche, \\
31028 Toulouse, France}

\maketitle\abstracts{ The concept of coded mask imaging in theory
and in practice is reviewed, with particular emphasis on image
reconstruction techniques.  The techniques are simple in principle
but  become more complicated when one takes into account real,
`as-built', instruments, as opposed to idealised imaginary ones.
Procedures are discussed with particular reference to the
instruments of Integral. }

\section{Introduction}

A coded mask telescope can be regarded as an imaging system in
which the `image' is so blurred  that the point source response
function (PSF) often extends over the whole of the detector plane.
The `image' referred to here, and the PSF  which describes it
($P_T$), are simply a shadow of the mask, or of part of it. Image
processing procedures can of course reduce or remove blurring and
this is a what is done by the various image reconstruction
algorithms which are used with coded mask telescopes. After
processing by such an algorithm one has a more normal response
function -- pyramidal or Gaussian for example. It is this
post-processing PSF ($P_{PP}$) which is normally quoted for a
coded mask telescope. $P_{PP}$  is really the response of a
pseudo-telescope which consists of a hardware component (often on
a satellite) and a software component (usually on the ground).

Image processing techniques can de-blur an image, but even in
conventional systems where the blurring is limited, they do so at
the expense of a degradation in the noise level. In the Fourier
domain, which provides a convenient way of removing the effects of
a PSF which is spatially independent, the degradation results from
the necessity of multiplying certain frequency components by large
factors. For de-blurring algorithms based on matrix techniques,
large terms in the inverse matrix play the same r\^ole.

We shall see below that, at least in principle, there are designs
of coded mask systems which avoid any such selective amplification
of Fourier terms and for which the inverse matrix has no
exceptionally large elements. However, one still finds that each
point in the image is subject to noise from most or all of the
detector plane. This is an intrinsic problem of non-focussing
instruments. In the absence of a mirror or lens to concentrate the
flux, the signal must be collected from the whole of the detector
and the background is harvested with the signal. % - the weeds
% with the wheat.

Despite having worked with coded masks telescopes for 25 years, it
is the view of the author that, if at all possible, one should
avoid their use! Systems which concentrate the flux will normally
be preferable.

 The fact remains though, that for certain X-ray and
gamma-ray imaging applications coded mask telescopes offer the
best available solution. The most important alternatives are
grazing incidence optics, tracking detectors (Egret, Glast ...)
and Compton techniques. The first two are limited to low energies
($<$100 keV) and to high energies ($>$50 MeV) respectively.
Compton techniques have a limited effective area (important when
one is photon-limited - for burst studies, for example) and can
never offer the very best energy resolution because the
uncertainties in two or more detectors are involved. Even within
the energy  band for which narrow field instruments can use
focussing optics, the wide field imaging capabilities of a coded
mask system are often invaluable (e.g. the Beppo-Sax WFC).

\section{Image reconstruction techniques}

\subsection{Correlation techniques}

A method of image reconstruction widely used with coded mask
telescopes is  correlation of the recorded data pattern with some
function $F$ which is a representation $P_T$ - {\it i.e.} of a
shadow of the mask pattern. It can be shown \cite{skinnercapri}
that if  $F$ has an appropriate scaling and offset, and if the
observation is dominated by Gaussian noise on a uniform
background, then this yields a minimum error estimate of the
intensity of a single point source.

For this reason, the correlation image offers the highest
sensitivity to a single point source; but it is not necessarily
the best method for imaging a complex field. If $P_T$ is position
independent, the post-processing PSF, $P_{PP}$, obtained in this
way is the autocorrelation function of the mask pattern (with an
offset and scaling). This is guaranteed to have a maximum value at
zero shift, corresponding to the true source position. It may,
however, have side peaks and spurious responses elsewhere
(`side-lobes', or `ghosts').

In practice for any reasonable mask pattern (an even for some less
reasonable ones! \cite{skinnercapri}) the correlation approach
yields a robust, if imperfect, image.

Reconstruction by correlation is equivalent to multiplying a
vector of observations (the intensities in detector pixels) by a
reconstruction matrix $R$ in which the rows represent the
different shifts of the mask pattern. But in cases where $P_T$ is
position independent (and preferably cyclic), Fast Fourier
Transform (FFT) techniques can be used to obtain the correlation
image very efficiently. For certain mask designs, reconstruction
by Hadamard or other transforms can be even faster than by FFT.

Working via the Fourier domain in principle allows compensation to
be made for different coding efficiencies at different spatial
frequencies. Wiener filtering \cite{wiener} does this with an
optimum trade-off between errors due to residual side-lobes on the
one hand and those due to noise-enhancement on the other.

\subsection{Back Projection}

A method of image reconstruction, useful when the number of events
is small, is to project back from the detected position of each
photon onto a map of the sky and to increment those pixels which
are consistent with the origin of that photon. This is equivalent
to forming an image by correlation with a 0/1 representation of
the mask pattern.

\subsection{Matrix Methods}

If the autocorrelation of the coding pattern is not bi-valued with
a single central peak and flat wings, then ghosts of bright
sources will occur in correlation images.  They can in principle
be avoided by noting that at each of the $N_{D}$ points in the
detector one measures a linear combination of the contributions
from each point in the sky (plus background). Provided the problem
is not under determined, the inverse of the matrix representing
this coding can be used to obtain the sky distribution which led
to the observed data.

This process is equivalent to solving a set of linear equations
for the  intensities in $N_{Sky}$ sky pixels plus some number
$N_{BG}$ of parameter describing the detector background
($N_{BG}=1$ for a uniform but unknown background). Provided that
among the
 $N_D$  measurements, there are at least $N_D'$ which are linearly
independent of each other and if  $N_D' = N_{Sky}+N_{BG}$,  this
is possible.

If $N_D' < N_{Sky}+N_{BG}$ more measurements are needed to obtain
a unique solution. One approach is to make further observations
with different mask patterns \cite{antimask1,antimask2}. Another,
which is critically important for instruments like Integral-SPI
with few detector elements, is to make, say, $n$ observations with
different pointings in the same general direction - so-called
`dithering'. One then has $n N_d$ observations.  If the source
fluxes and background parameters are assumed constant, the number
of unknowns does not increase. Even if it is supposed that they
vary according to some simple model, then one may still have
gained.

If $N_D'$ (or $nN_D'$ with dithering) is greater than
$N_{Sky}+N_{BG}$, then one must use the Moore-Penrose generalised
inverse, which provides the best-fit solution to an
over-determined set of equations.

As is well-known, inverse matrix methods tend to be unstable.
Techniques are available \cite{rideout} which find a stabilised
reconstruction matrix $R$ which is optimum according to the same
Wiener criterion that is used in Fourier space. This stabilised
matrix is a compromise between the $R$ matrix which corresponds to
correlation imaging, which is best in the case of poor signal to
noise ratio where side lobes are of secondary importance, and the
inverse matrix with its perfect imaging but potentially disastrous
noise amplification.

\subsection{Non-linear techniques }

All of the above techniques lead to an intensity estimate in each
image pixel which is a linear combination of the data. When the
number of detected events per pixel is small the statistics are
not strictly Gaussian but Poissonian.  Correlation involves
combining many pixels, with both positive and negative multipliers
and the differences tend to be small, but it can be argued that it
is better to use non-linear methods which take the Poissonian
nature of the noise into account.

There are other reasons for considering techniques which are not
in the above sense linear. These reasons vary from the aesthetic
to the pragmatic - from the appeal of the clearly defined
assumptions of a Bayesian approach, via a preference for solutions
without un-physical negative intensities,  to the fact that
certain techniques just `seem to work'. For lack of space,
non-linear techniques will only be briefly mentioned here and not
be reviewed in any detail.

The Poissonian nature of the statistics can be taken into account
by using Maximum Likelihood (ML) techniques
\cite{skinner_nottingham}. ML naturally leads to a constraint that
the intensity estimates shall be positive, but a non-negativity
constraint can be imposed while adopting other optimisation
parameters \cite{positivity}. Normally no assumption is made about
the relative probability of different images, {\it i.e.} no prior
probability function is used.

Maximum Entropy is a technique which seems to be well suited to
coded mask image reconstruction, for which it has been widely
discussed \cite{maxent1,maxent2}. It is an iterative Bayesian
method, sililar to maximum likelihood, but with a prior
probability function (the entropy) which favours flatter images.
Other iterative techniques, such `direct deconvolution' \cite{li},
have been proposed.

Certain techniques try to find a model, consistent with the data
and comprising a limited number of components. Either the Pixon
method \cite{pixon} or Wavelet techniques \cite{wavelet} could be
used in this way. Often the simplest description of the field (and
hence in a sense the Maximum Entropy solution) is just a list of
point sources. Supposing pure point source model leads to IROS
(\S\ref{iros}).

\section{Idealised (`imaginary') coded mask telescopes}

\label{sect_imag} Usually the mask pattern consists of pixels
which are of equal size and which are either transparent or
opaque, though the reconstruction methods discussed above are
mostly not limited to such cases.  In comparing alternative
designs, one may use the concept of `coding power'
\cite{skinnercapri}, which is a measure of the background-limited
point source sensitivity obtainable with correlation techniques.

The coding power  is essentially the root-mean-square deviation of
the transmission of the mask about its mean value. As the
transmission is limited to the range $0$--$1$ one can quickly
deduce that the highest value of the coding power, and hence the
best point-source sensitivity in the detector background dominated
case is obtained (a) if only the extreme values $0,1$ occur and
(b) if the mean transparency is 50\%. If  sky-related events such
as cosmic diffuse background and source counts make a significant
contribution to the noise, then (a) remains true, though the
optimum transparency may change \cite{xx,yy,zz}.

To avoid side lobes in correlation images (and to minimise the
interdependence of image pixels irrespective of the reconstruction
method) one should choose a mask pattern whose autocorrelation
function is flat away from the central peak. If the coding can be
contrived to be cyclic, this is possible by using patterns related
to cyclic difference sets \cite{g_and_p,ura,psw}.  As well as
exhibiting no structure away from the central peak in the PSF in a
correlation image, these patterns have $0/1$ transmission and can
have very close to 50\% transparency.  So they have optimum
sensitivity.

For such mask designs {\bf all of the above mentioned linear
reconstruction methods are equivalent} \footnote {Here terms which
are of the order $1/N$, where there are $N$ elements in the mask,
and hence usually very small, are ignored.}.
 Thus in principle one has a way of designing a coded mask system
which is as sensitive as possible and multiple  ways of
reconstructing the image, any one of which has imaging properties
which are as good as possible.

\section{Real systems}

So much for the world of Dr. Pangloss \footnote{\,``Dans ce
meilleur des mondes possibles ... tout est au mieux."
% All is for the best in the best of all possible worlds
 Voltaire, {\it Candide,} 1759.}.
 We can now proceed to strip off the various
idealisations which lead to this fictional situation.

\subsection{Detector sampling}
Often in the literature it is implicitly assumed that the sources
conveniently lie in directions such that the shadows of the mask
elements are coincident with detector pixels which are equal
either to the size of the mask elements or to a submultiple of
that size. In general this will not be the case and a blurred form
of the mask shadow will be recorded. Even if the detector readout
is continuous, the recorded pattern will still generally be
blurred, because the spatial resolution will not be infinitely
good.

Provided that the detector response is independent of the position
within the detector (except perhaps for binning associated with
pixels which are some submultiple of the mask pixel size), then
the same mask patterns continue to offer optimal properties. But
in these circumstances it is the coding power of the blurred mask
which matters and the sensitivity is reduced.

\subsection{Cyclic coding}
It is not always possible to ensure that the coding is cyclic; in
fact it is not always desirable. In principle, cyclic coding can
be achieved by having a mask which contains several repeats of a
basic pattern and a detector which registers one cycle. But unless
a fine-structured collimator covering the whole detector is used
to reduce to zero the response outside a defined region, there
will always be some source directions for which a partial shadow
is recorded. Such collimators can be used but they attenuate the
signal even within the `fully-coded' region and the loss in
sensitivity may be more important than the improvement in imaging
properties. .

 If the coding is not truly cyclic, the
optimum patterns mentioned in \S\ref{sect_imag} are still
sometimes selected ({\it e.g.} TTM, Integral ...) because they are
well defined and are usually not {\bf worse} than a random
pattern. But when the recorded data does not correspond to a whole
number of cycles of the the pattern shadow, they rapidly cease to
be any {\bf better} than a random pattern.

\subsection{Other effects}
Other respects in which real systems may differ from idealised
ones are listed in Table 1. These effects lead to a raw PSF $P_T$
that differs greatly from the idealised response, which would just
be a translation of a perfect mask pattern. What is worse, is that
the differences tend to be a function of source position.

\begin{centering}

\begin{table}
\caption[]{Imperfections which can arise in non-ideal systems}
\vspace{2mm} \hspace{10mm} \scriptsize{
\begin{tabular}{|ll|}
 \hline
 \multicolumn{1}{|c}{\underline {Mask}} & \\
Non-cyclic   &\multicolumn{1}{c|}{ \underline {Detector}}\\
Closed element : imperfect absorption
  & Detector finite position resolution\\
Open element : imperfect transmission  &Detector efficiency non-uniformities \\
Effects of mask finite thickness & Detector response dependent on off-axis angle\\
Obstructions in  the mask plane  &Detector background non-uniform \\
                                  & Gaps in the detector plane\\
  \multicolumn{1}{|c} {\underline {Other } } &Dead/inactive pixels in the detector plane\\
  \multicolumn{1}{|l} {Shielding (collimation) imperfect} ~ ~ ~ ~ ~ ~ ~  &\\
   \multicolumn{2}{|l|}  {Obstructions between the detector and the mask}\\
   \multicolumn{2}{|l|}  {Pointing errors; Pointing drift }\\
  \multicolumn{2}{|l|}   {Leaks onto the detector from far outside the field of view }\\
\hline
\end{tabular}
}
\end{table}
\end{centering}

\section{Real life data analysis}

\label{iros}

In principle a detailed knowledge of the instrument allows the
coding matrix element for each detector pixel and each possible
source direction to be determined. In theory, any of the more
general of the methods outlined above (those which do not assume a
position independent response function) can then deal with all of
the effects in Table 1.

The matrices  describing the response are of dimensions
$(N_{Sky}+N_{BG}) \times nN_D'$ and so could potentially have up
to $10^{12}$ elements for images of the order of $1000\times 1000$
pixels. Techniques using detailed response matrices, including use
of the stabilised pseudo-inverse matrix, are feasible for low
resolution instruments with a limited number of detector pixels.
The software for the SPI spectrometer of Integral, with 19
detector elements, depends heavily on this approach. But using
such large matrices for the production of an image becomes
impractical when the number of detector pixels and of resolvable
sky pixels is large. For Integral, the  ISGRI and  PICSIT
detectors of IBIS have 4096 and 16384  detector pixels
respectively. With  JEM-X the number rises to $>$50000.

Iterative Removal Of Sources  (IROS) \cite{iros} takes advantage
of the fact that with systems with a large number of pixels the
sensitivity to extended sources is poor. Consequently one is
usually dealing with the problem of finding a limited number of
point sources and determining their positions and intensities.

With IROS, the first stage is to obtain an image which is good
enough to find the approximate position of the brightest source.
The image does not need to be free of artifacts provided the most
significant point corresponds to a real source. Typically
approximations and short-cuts are taken by filling in missing data
with local averages and by making simplified corrections for
non-uniformities and deficiencies. Generally Fourier techniques
have to be used and these assume a position-independent response
function, even though this may not strictly be the case. For
smaller numbers of pixels the position-dependent response may be
taken into account, but a comparatively coarse image sampling may
be used to reduce the problem size.

Having identified the strongest source in the field, the position
and intensity of that source (and the values of any unknown
background parameters) are optimised by fitting {\bf in the data
space.} At this stage all the known imperfections in the coding
and in the instrument are taken into account. After subtracting
the data predicted for the fitted source, an image reconstruction
based on the residual data is used to search for further sources.
If more sources are found they are in turn subtracted, but
importantly the parameters of all of the sources are re-optimised
each iteration.

\section{Conclusions}

The fact that the source intensity will be a function of energy
and perhaps of time has not been considered here. The above
principles can be applied to successive subsets of the data in
energy channel or time. In the former case, off-diagonal terms in
the energy response matrix describing the entire hardware plus
software `instrument' must be taken into account in interpreting
the spectra obtained.

It has here been possible discuss only a few of the vast range of
different coded mask image reconstruction and data analysis
techniques that have been suggested,  and those only in outline.
The objective has been review some of the key issues ahead of the
Integral mission which will place into orbit no fewer than four
telescopes using the coded mask technique. Experience with data
from Integral will no doubt lead to the development of even more
sophisticated and effective analysis techniques.

\end{document}